\documentclass[useAMS, usenatbib]{mnras}
\usepackage{epsfig}
\usepackage{amsfonts}
\usepackage[usenames]{color}
\usepackage[usenames,dvipsnames]{xcolor}
\usepackage{url}
\usepackage{subfigure}
\usepackage[english]{babel}
\usepackage{times,graphicx,amsmath,amssymb,aas_macros,epstopdf,hyperref}
\usepackage[normalem]{ulem}
\usepackage[T1]{fontenc} 
\usepackage{ae,aecompl} 
\usepackage{multirow}
\usepackage{xspace} 
\usepackage{units}
\usepackage{wasysym}
\usepackage{newtxtext,newtxmath}
\usepackage{arydshln}

\usepackage{hyperref}
\hypersetup{
    colorlinks=true,
    linkcolor=blue,
    citecolor=blue,
}

\def\etal{{\frenchspacing\it et al.}}
\def\ie{{\frenchspacing\it i.e.}}

\def\etc{{\frenchspacing\it etc.}}

\def\be{\begin{equation}}
\def\ee{\end{equation}}
\def\ba{\begin{eqnarray}}
\def\ea{\end{eqnarray}}

\def\nn{\nonumber}

\def\LaTeX{L\kern-.36em\raise.3ex\hbox{a}\kern-.15em
    T\kern-.1667em\lower.7ex\hbox{E}\kern-.125emX}

\begin{document}

\voffset-1.25cm
\title[A tomographic BAO and RSD analysis in Fourier space]{The clustering of galaxies in the completed SDSS-III Baryon Oscillation Spectroscopic Survey: a tomographic measurement of structure growth and expansion rate from anisotropic galaxy clustering in Fourier space}
\author[Zheng, Zhao, Li \etal]{
\parbox{\textwidth}{
Jinglan Zheng$^{1,2}$\thanks{\url{jlzheng@nao.cas.cn}}, Gong-Bo Zhao$^{1,3,4}$\thanks{\url{gbzhao@nao.cas.cn}}, Jian Li$^{1,2}$, Yuting Wang$^{1}$, Chia-Hsun Chuang$^{5,6}$, Francisco-Shu Kitaura$^{7,8}$, Sergio Rodriguez-Torres$^{9,10}$}
\vspace*{15pt} \\
$^{1}$ National Astronomy Observatories, Chinese Academy of Science, Beijing, 100101, P. R. China \\
$^{2}$ University of Chinese Academy of Science, Beijing, 100049, P. R. China \\
$^{3}$ College of Astronomy and Space Sciences, University of Chinese Academy of Sciences, Beijing 100049, China \\
$^{4}$ Institute of Cosmology \& Gravitation, University of Portsmouth, Dennis Sciama Building, Portsmouth, PO1 3FX, UK \\
$^{5}$  Kavli Institute of Particle Astrophysics and Cosmology \& Physics Department, Stanford University, Stanford, CA 94305, USA \\
$^{6}$ Leibniz-Institut f\"ur Astrophysik Potsdam (AIP), An der Sternwarte 16, D-14482 Potsdam, Germany  \\
$^{7}$ Instituto de Astrof\'{\i}sica de Canarias (IAC), C/V\'{\i}a L\'actea, s/n, E-38200, La Laguna, Tenerife, Spain \\
$^{8}$ Departamento Astrof\'{\i}sica, Universidad de La Laguna (ULL), E-38206 La Laguna, Tenerife, Spain \\
$^{9}$ Departamento de F\'isica Te\'orica M8, Universidad Aut\'onoma de Madrid (UAM), Cantoblanco, E-28049, Madrid, Spain\\
$^{10}$ Instituto de F\'isica Te\'orica, (UAM/CSIC), Universidad Aut\'onoma de Madrid, Cantoblanco, E-28049 Madrid, Spain\\
}
\date{\today} 
\pagerange{\pageref{firstpage}--\pageref{lastpage}}

\label{firstpage}

\maketitle

\begin{abstract} 
We perform a tomographic structure growth and expansion rate analysis using the monopole, quadrupole and hexadecapole of the redshift-space galaxy power spectrum derived from the Sloan Digital Sky Survey (SDSS-III) Baryon Oscillation Spectroscopic Survey (BOSS) Data Release 12 combined sample, which covers the redshift range of $0.20<z<0.75$. By allowing for overlap between neighbouring redshift slices in order to extract information on the light-cone, we successfully obtain joint BAO and RSD constraints with a precision of $2-3\%$ for $D_A$, $3-10\%$ for $H$ and $9-12\%$ for $f\sigma_8$ with a redshift resolution of $\Delta z\sim0.04$.  Our measurement is consistent with that presented in \citet{WangRSD}, where the analysis is performed in configuration space. We apply our measurement to constrain the $f(R)$ gravity model, and find that the 95\% CL upper limit of ${\rm log_{10}}B_0$ can be reduced by 11\% by our tomographic BAO and RSD measurement. We make our joint BAO and RSD measurement publicly available at \url{https://github.com/Alice-Zheng/RSD-data}

\end{abstract}


\section{Introduction}
\label{sec:intro}

Physics behind the accelerating expansion of the Universe, which was discovered in 1998 \citep{Riess1998,Perlmutter98}, remains unveiled. In principle, introducing dark energy (DE) as a dominating energy component of the Universe at the current epoch (see \citealt{DEreview} for a recent review on dark energy), or extending Einstein's general relatively on cosmological scales, dubbed the modified gravity (MG) scenario (see \citealt{MGreview} for a recent review on modified gravity), can be possible origins of the cosmic acceleration. Although these two scenarios can be degenerate at the level of the cosmic expansion, they are distinguishable at the level of structure formation. Observationally, large spectroscopic galaxy surveys offer key probes for both DE and MG through measurements of specific three-dimensional patterns of galaxy clustering including the Baryonic Acoustic Oscillations (BAO) and Redshift Space Distortions (RSD). 

The observable of BAO is the excess in galaxy clustering at about 150 $\,\rm Mpc$ on the comoving scale, which is due to the interaction between photons and baryons in the early universe \citep{Peebles1970}. The BAO scale can be derived from two-point correlation functions or power spectra multipoles of galaxies in redshift space. By virtue of the Alcock-Paczynski (AP) effect \citep{AP1979}, which quantifies a difference of the BAO distance scales between radial and transverse directions due to an improper choice of the `fiducial' cosmology to convert redshifts to distances, the Hubble parameter and the angular diameter distance at an effective redshift $z$, $H(z)$ and $D_A(z)$ respectively, can be estimated from the anisotropic galaxy clustering. As the BAO scale is largely immune to systematics \citep{Ross2011}, it has been widely used as a standard ruler to probe the expansion history of the Universe, facilitating a power tool for the study of DE.

RSD produce another kind of anisotropy in the galaxy clustering. Unlike the AP effect, RSD are due to peculiar motions of galaxies affected by local gravitational potentials. As the galaxy clustering is observed in redshift space, in which the peculiar motions only alter the clustering along the line of sight, an anisotropy in the galaxy clustering is produced, which is directly related to gravity. In the linear regime, RSD yield an angle-dependent boost of the power spectrum amplitude by a factor of $(1 + \beta\mu^2)^2$~\citep{KaiserN1987}, where $\mu$ is the cosine of the angle between the galaxy pair and the line-of-sight vectors, and $\beta \equiv f/b$, the ratio between the growth rate $f$ and the galaxy bias $b$. This provides a direct measure of the growth rate of the Universe, which is one of the key probes of gravity on cosmological scales.

Joint measurements of BAO and RSD can in principle break the `dark degeneracy' between DE and MG, which is key to understand the cosmic acceleration. In order to maximise the BAO and RSD information extracted from the survey volume, methods of tomographic analysis on the lightcone have been developing, including the overlapping redshift slicing method \citep{GB2017,WangRSD,Wangtomo16} and optimal redshift weighting schemes \citep{ZPW,zwBAOmock,zwRSD,RR18,DD18,Zhao18,Zhu18}. As demonstrated in \citet{NAZhao17} and \citet{Zhao18}, these methods can effectively extract additional information on the lightcone, which generically improves constraints on dark energy and modified gravity. The optimal redshift weighting method is more computationally efficient, as it allows for measuring the linear combinations of power spectra at various redshifts, which are most sensitive to the cosmological parameters, without subdividing the galaxy sample. However, it requires a robust modelling of the time evolution of key cosmological quantities including the BAO, RSD, the bias, \etc \ in the first place, which can be theoretically challenging. On the other hand, the overlapping redshift slicing method is more computationally expensive, but it does not require an assumption of the temporal evolution of the system, which is less subject to theoretical systematics.

In this work, we apply the overlapping redshift slicing method developed in \citet{GB2017,WangRSD,Wangtomo16} to the BAO and RSD analysis in Fourier space using the BOSS DR12 galaxy catalog, and make cosmological implications. This paper is structured as follows. In Section \ref{sec:method}, we describe the methodology in our analysis, including the galaxy and mock catalogs used for this analysis, the theoretical template and details for parameter estimation. In Section \ref{sec:results}, we show the main results including the mock tests, tomographic measurements of BAO and RSD parameters, as well as a cosmological implication on observational constraints on the $f(R)$ gravity. The last section is devoted to conclusion and discussions.

\section{Methodology}
\label{sec:method}

In this section, we describe the data used in this analysis (including the galaxy and mock catalogs), the theoretical template for the joint BAO and RSD measurement, and the method used for parameter estimation.

\subsection{The Data}

The observational dataset used in this work is obtained by the Sloan Digital Sky Survey (SDSS-III) Baryon Oscillation Spectroscopic Survey (BOSS). Using a $2.5$ metre-aperture Sloan Foundation Telescope \citep{SDSStelescope} at the Apache Point Observatory, the BOSS program covers around 10, 000 square degrees of the sky. The BOSS team has obtained spectra of over 1.5 million galaxies brighter than $i=19.9$ and approximately 170, 000 new quasars in the redshift range of $z\in[2.1,3.5]$. The spectrograph, filter and pipeline of BOSS are described in \citet{ugrizFilter,BOSSpipeline,BOSSspectro}. 

The galaxy catalogue for this analysis is built upon the BOSS Data Release (DR) 12 combined sample, which is a coherent combination of two distinct targets of LOWZ and CMASS. The stellar-mass incompleteness of the LOWZ and CMASS samples was discussed in \citet{Leauthaud16}, and its impact on the clustering was studied in \citet{Saito16} and \citet{Rodriguez-Torres16}. The DR12 combined catalogue was reduced from observations using the pipeline described in \citet{Reid16}, where the survey footprint, veto masks and survey systematics are taken into account when creating the data and random catalogues. The redshift range of this sample is $z\in[0.2,0.75]$, containing approximately $865, 000$ and $330, 000$ galaxies in the North Galactic Cap (NGC) and South Galactic Cap (SGC) respectively. 

For the purpose of estimating the data covariance matrix, and of validating our data analysis pipeline, mock galaxy catalogues are required. In this work, we use the MultiDark-Patchy (MD-Patchy) mock catalogue \citep{Kitaura2016}, which offers 2048 realisations of mock galaxy distributions, matching the spatial and redshift distribution of the DR12 combined sample. As demonstrated in \citet{Kitaura2016}, the Patchy mocks accurately reproduce the two-point and three-point statistics of galaxy clustering in the BOSS DR12 sample, which validates it for the use of likelihood analysis for the DR12 sample. The light-cone of the Patchy mocks was constructed using ten redshift slices, which permits the determination the time evolution of the galaxy bias, the growth, and the peculiar motion. We refer to \citet{Kitaura2016} for more details of the implementation of the light-cone construction.

The fiducial cosmology used in the analysis is the same as that used for producing the Patchy mocks, namely, 
\be\label{eq:fid}\left\{\Omega_{\rm M}, \Omega_{\rm b}, \Omega_{\rm K}, h, \sigma_8\right\} = \left\{0.307115, 0.0480, 0, 0.6777, 0.8288 \right\}\ee which is consistent with the results from the Planck collaboration \citep{Planck2015}.

\subsection{The overlapping redshift slicing}

To extract information from the past lightcone of the survey, we adopt the overlapping redshift slicing (ORS) method developed and applied in \citet{GB2017,WangRSD,Wangtomo16}. The essence of the ORS method is to subdivide the galaxy sample into numerous overlapping redshift slices, in order to guarantee that the number of galaxies in each redshift slice is sufficiently large to yield a decent BAO and RSD measurement, and that the redshift slicing is sufficiently fine so that the key tomographic information on the lightcone is extracted. In \citet{GB2017,WangRSD,Wangtomo16}, where the same galaxy and mock catalogues are used for BAO and/or RSD analyses (see Table \ref{tab:ORSanalysis} for details of these analyses), the DR12 combined sample was subdivided into nine overlapping redshift slices, as detailed in Table \ref{tab:zbins}. As detailed in \citet{GB2017}, our redshift-slicing scheme can well balance the redshift resolution and the complementarity of the information between overlapping bins, and an analysis of RSD in configuration space using the same redshift binning has demonstrated the effectiveness of our binning scheme \citep{WangRSD}. In this work, we adopt the same ORS for a joint BAO and RSD analysis in Fourier space, which is complementary to the joint BAO and RSD analysis in configuration space performed in \citet{WangRSD} \footnote{Note that although the correlation function and power spectra are directly related to each other by a Fourier transformation in the ideal case (\ie, a survey with an infinite volume), they are complementary for realistic galaxy surveys. On the other hand, \cite{WangRSD} only used the monopole and quadrupole, while we additionally use the hexadecapole in this analysis.}

\begin{table}
\begin{center}
\begin{tabular}{cc}
\hline\hline
Analysis                     & Reference \\
\hline
BAO in $s$-space     &  \citet{Wangtomo16}\\
BAO in $k$-space     &  \citet{GB2017}\\
BAO+RSD in $s$-space     &  \citet{WangRSD}\\
BAO+RSD in $k$-space     &  This work\\
\hline\hline
\end{tabular}
\end{center}
\caption{A list of tomographic BAO or RSD analyses using the overlapping redshift slicing shown in Table \ref{tab:zbins}.}
\label{tab:ORSanalysis}
\end{table}%

\begin{table}
\begin{center}
\begin{tabular}{ccc}
\hline\hline
   redshift bin index &  redshift range & effective $z$   \\ \hline

$z$ bin 1	&	$0.20	<	z	<	0.39$	&	0.31 \\ 
$z$ bin 2	&	$0.28	<	z	<	0.43$	&	0.36	\\
$z$ bin 3	&	$0.32	<	z	<	0.47$	&	0.40 \\
$z$ bin 4	&	$0.36	<	z	<	0.51$	&	0.44	\\
$z$ bin 5	&	$0.40	<	z	<	0.55$	&	0.48 \\
$z$ bin 6	&	$0.44	<	z	<	0.59$	&	0.52 \\
$z$ bin 7	&	$0.48        <	z	<	0.63$	&	0.56 \\
$z$ bin 8	&	$0.52        <	z	<	0.67$	&	0.59 \\
$z$ bin 9	&	$0.56        <	z	<	0.75$	&	0.64 \\

\hline\hline
\end{tabular}
\end{center}
\caption{The overlapping redshift slicing applied on the DR12 combined sample for analyses shown in Table \ref{tab:ORSanalysis}.}
\label{tab:zbins}
\end{table}

\subsection{Measurements of the power spectrum multipoles}

As we use the same redshift slicing scheme as that in \citet{GB2017} for this analysis, we use the power spectrum multipoles (up to the hexadecapole) measured in \citet{GB2017} using a Fast Fourier Transformations (FFTs) method \citep{FFTPk}. For the measurement, galaxies and random catalogues are placed in a cubic box with $L=5000 \ h^{-1}$ Mpc a side, which is divided into $1024^3$ cubic cells for calculating the over-density field, and for the FFTs, and we follow the prescription developed in \citet{Jing05} to correct for the aliasing effect of the FFTs.

\subsection{The template}

\subsubsection{Modelling the power spectrum in redshift space}
We use the extended TNS (eTNS) prescription \citep{TNS2010} to model the anisotropic power spectrum in redshift space, as implemented for the RSD analyses in Fourier space using datasets of BOSS DR12 \citep{Beutler2017} and eBOSS DR14 \citep{Zhao18,HGM18}.

The eTNS model reads,
\begin{equation}
 \begin{split}
P_{\rm g}(k,\mu) &= e^{-(fk\mu\sigma_v)^2}\left[P_{{\rm g},\delta\delta}(k)\right. \\
&\;\;\; + 2f\mu^2P_{{\rm g},\delta\theta}(k) + f^2\mu^4P_{\theta\theta}(k) \\
&\;\;\; + b_1^3\sum_{m,n=1}^3\, \mu^{2m}\,\beta^{n} A_{mn} \\
&\;\;\;+ b_1^4\sum ^{4}_{n=1}\sum ^{2}_{a,b=1}\mu ^{2n}\left( -\beta\right) ^{a+b}\left.B^{n}_{ab}\right]
 \label{eq:tns}
 \end{split}
\end{equation}
where  $f\equiv \frac {d\ln D\left( a\right) }{d\ln a}$ denotes the logarithmic growth rate, $\mu$ is the cosine of the angle between the wavenumber vector $k$ and the line-of-sight direction, and $\sigma_{v}$ is treated as a free parameter to be marginalised over.

The overall exponential damping factor in Eq (\ref{eq:tns}) encodes the Fingers-of-God (FoG) effect. The terms in the square brackets extend the linear Kaiser model, where the first three terms are auto- and cross-power spectra of the matter density field $\delta$ and of the divergence of the peculiar velocity field $\theta$, while the 
the last two (the A and B) terms correct for higher-order correlations. We compute the power spectra terms using the regularised perturbation theory (RegPT) up to second order \citep{RegPT} \footnote{Available at \url{http://www2.yukawa.kyoto-u.ac.jp/~atsushi.taruya/regpt_code.html}}, and calculate the A and B terms using the standard perturbation theory (SPT) \citep{TNS2010}. Note that all these terms except for $P_{\theta\theta}(k)$ contain the the linear bias $b_{1}$ and the local non-linear bias $b_{2}$ \footnote{Other bias terms $b_{s2}$, $b_{\rm 3nl}$ can be reduced to $b_{1}$ and $b_{2}$ terms following \citet{C-S-S-2012}.}.

\subsubsection{The Alcock-Paczynski effect}

The Alcock-Paczynski (AP) effect \citep{AP1979} changes the galaxy clustering due to the incorrect input cosmology to convert redshifts to distances. Mathematically, it alters the anisotropic galaxy clustering shown in Eq (\ref{eq:tns}) in the following way \citep{Ballinger1996},
\ba P_g(k,\mu) \rightarrow P_g(k',\mu'), \ea where 
\ba
&&k' = \frac{k}{\alpha_{\perp}}\left[1 + \mu^2\left(\frac{1}{F^2} - 1\right)\right]^{1/2},
\label{eq:scaling1} \nn \\
&&\mu' = \frac{\mu}{F_{}}\left[1 + \mu^2\left(\frac{1}{F^2} - 1\right)\right]^{-1/2}, \nn \\
&&F = \alpha_{\parallel}/\alpha_{\perp}.
\label{eq:scaling2}
\ea

Then the power spectrum multipoles can be calculated as,
\begin{align}
P_{\rm \ell}(k) &= \left(\frac{r_s^{\rm fid}}{r_s}\right)^3 \frac{(2\ell + 1)}{2\alpha^2_{\perp}\alpha_{\parallel}}\int^1_{-1}d\mu\; P_{\rm g}\left(k', \mu'\right)\mathcal{L}_{\ell}(\mu),
\label{eq:multi}
\end{align}
where $r_s$ denotes the sound horizon at the recombination. The factor $\left(\frac{r_s^{\rm fid}}{r_s}\right)^3\frac{1}{2\alpha^2_{\perp}\alpha_{\parallel}}$ accounts for the difference in the cosmic volume in different cosmologies \citep{Beutler2017a}. 

\subsubsection{The survey window function}

Due to the irregularity and the finite size of the survey volume of the BOSS survey, the observed power spectrum is the theoretical power spectrum convolved with the window function. We follow the method developed in \citet{Wilson2015}, which is an efficient way to reduce the three-dimensional convolutions to one-dimensional Hankel transformations, which can be done rapidly using the FFTlog algorithm \citep{Hamilton2000}. We refer to \citet{GB2017} for more details on the convolution of the survey window function for the ORS used in this work, and for details on how to convolve the window function of the survey with the model; briefly we  follow the method developed in \citet{windowpk} and make use of FFTlog libraries \citep{FFTlog}. The window functions used for this analysis are identical to those shown in Fig. 9 in \citet{GB2017}.

\subsection{The parameter estimation}

As the target selection is slightly different for the North Galactic Cap (NGC) and in the 
South Galactic Cap (SGC),\footnote{The density for some chunks (2-6) in the North Galactic Cap is generally lower compared to the rest of the dataset, therefore the expected bias parameters are different for NGC and SGC.} We treat the two areas separately \citep{Beutler2017,GB2017}, which leaves eleven free parameters to be determined for each redshift slice.

For a given set of parameters, we use a modified version of {\tt CAMB}  \footnote{Available at \url{http://camb.info}} \citep{camb} to compute the theoretical prediction, and then use {\tt CosmoMC} \footnote{Available at \url{https://cosmologist.info/cosmomc/}} \citep{cosmomc} to sample the parameter space using the Monte Carlo Markov Chain (MCMC) method. The likelihood function to be maximised by {\tt CosmoMC} is,

\ba
 \chi^2 (\bold{p}) \equiv  \sum_{i,j}^{\ell,\ell'}  \left[P^{\rm d}_{\ell} (k_i, \bold{p}) -P^{\rm th}_{\ell}(k_i) \right] F^{\ell,\ell'}_{ij} \left[P_{\ell'}^{\rm d}(k_j, \bold{p}) -P^{\rm th}_{\ell'}(k_j)\right] \nonumber
\ea
where superscripts $^{\rm d}$ and $^{\rm th}$ denote data and theoretical prediction respectively, ${\bf p}$ stands for a collection of parameters shown in Table \ref{tab:param}, and $F^{\ell,\ell'}_{ij}$ is the inverse of the data covariance matrix estimated from the Patchy mock catalogues. We use the power spectrum measurement in the wavenumber range of $k\in[0.015,0.15] h \ {\rm Mpc^{-1}}$ to avoid contaminant from both observational and theoretical systematics \citep{Beutler2017}. We follow \citet{Will2014} to perform a rescaling of the uncertainty of each parameter returned by MCMC, to correct for the fact that finite number of mocks are used to estimate the data covariance matrix.

\begin{table}
\begin{center}
\begin{tabular}{ccc}
\hline\hline
Parameter & Meaning &Prior \\
\hline
$\alpha_\parallel$ & The radial BAO dilation parameter& [0.8,1.2] \\
$\alpha_\perp$ & The transverse BAO dilation parameter& [0.8,1.2] \\
$f\sigma_8$ & The RSD parameter & [0, 1] \\
$b_1^{\rm NGC}\sigma_8$ &The linear bias for the NGC & [0.5,2.1] \\
$b_2^{\rm NGC}\sigma_8$ &The nonlocal bias for the NGC & [0,4] \\
$\sigma_v^{\rm NGC}$ & The velocity dispersion for the NGC & [1,9] \\ 
$N^{\rm NGC}$ & The correction to the shot noise for NGC & [-2000,2000] \\
$b_1^{\rm SGC}\sigma_8$ &The linear bias for the SGC & [0.5,2.1] \\
$b_2^{\rm SGC}\sigma_8$ &The nonlocal bias for the SGC & [0,4] \\
$\sigma_v^{\rm SGC}$ & The velocity dispersion for the SGC & [1,9] \\ 
$N^{\rm SGC}$ & The correction to the shot noise for SGC & [-2000,2000] \\
\hline\hline
\end{tabular}
\end{center}
\caption{The free parameters, physical meaning and the flat priors used in the MCMC analysis for each redshift slice.}
\label{tab:param}
\end{table}%

\section{Results}
\label{sec:results}

\begin{table*}
\caption{Measurements (mean with the 68\% CL uncertainty) of BAO and RSD parameters including $\alpha_{\perp}$, $\alpha_{\parallel}$ and $f\sigma_8$ using $P_0+P_2$ (left part of the table) and $P_0+P_2+P_4$ (right) derived from the mock catalogues at nine effective redshifts. As a mock test, the absolute values of differences between the measurement and the expected values are shown. All measurements are multiplied by a factor of $100$ for illustration.}
\begin{center} 
\begin{tabular}{ccccccc}
\hline\hline 
        & &   Mock catalogue  $\left(P_0+P_2\right)$         &                    &    & Mock catalogue  $\left(P_0+P_2+P_4\right)$ &\\ \hline
$z_{\rm eff}$  & $\Delta\alpha_{\parallel}$ & $\Delta\alpha_{\perp}$&   $\Delta f\sigma_8$ & $\Delta\alpha_{\parallel}$ &  $\Delta\alpha_{\perp}$ &  $\Delta f\sigma_8$  \\ \hline
0.31	&	$	0.10   \pm	3.93  	$	&	$	0.48   \pm	3.56  	$	&	$	2.71    \pm  6.82   $	&	$	0.99   \pm	3.73  	$	&	$	0.48  	\pm	3.41  	$	&	$	1.59   \pm 5.77   	$	\\
0.36	&	$	0.68   \pm	3.70  	$	&	$	0.20   \pm	3.29  	$	&	$	3.73    \pm  7.48   $	&	$	0.86   \pm	3.30  	$	&	$	0.31   \pm	2.97  	$	&	$	1.67   \pm 5.42   	$	\\
0.40	&	$	0.58   \pm	3.56  	$	&	$	1.52   \pm	3.08  	$	&   $	1.73    \pm  6.42   $	&	$	0.28   \pm	3.02  	$	&	$	0.24  	\pm	2.68  	$	&	$	1.42   \pm 5.36   	$	\\
0.44	&	$	0.89   \pm	3.70  	$	&	$	1.74   \pm	3.07  	$	&	$	1.63    \pm  5.28   $	&	$	0.48   \pm	2.94  	$	&	$	0.26  	\pm	2.58  	$	&	$	0.21   \pm 5.12   	$	\\
0.48	&	$	1.04   \pm	3.67  	$	&	$	0.87   \pm	2.73  	$	&	$	0.05    \pm  6.86   $	&	$	0.71   \pm	2.72  	$	&	$	0.56  	\pm	2.35  	$	&	$	0.25   \pm 4.85   	$	\\
0.52	&	$	0.85   \pm	2.53  	$	&	$	0.73   \pm	2.31  	$	&	$	0.55    \pm  5.35   $	&	$	0.82   \pm	2.52  	$	&	$	0.21  	\pm	2.17  	$	&	$	0.57   \pm 4.48   	$	\\
0.56	&	$	0.64   \pm	3.80  	$	&	$	1.81   \pm	2.99  	$	&	$	3.61    \pm  6.34   $	&	$	0.95   \pm	2.42  	$	&	$	0.26  	\pm	2.09  	$	&	$	0.58   \pm 4.24   	$	\\
0.59	&	$	0.73   \pm	3.78  	$	&	$	0.92   \pm	2.82  	$	&	$	1.89    \pm  5.23   $	&	$	0.11   \pm	2.34  	$	&	$	0.39  	\pm	2.00  	$	&	$	0.91   \pm 4.24   	$	\\
0.64	&	$	1.67   \pm	3.63  	$	&	$	0.82   \pm	2.79  	$	&	$	3.92    \pm  7.43   $	&	$	0.15   \pm	2.27  	$	&	$	0.70  	\pm	1.90  	$	&	$	0.43   \pm 4.40   	$	\\

\hline\hline
\end{tabular}
\end{center}
\label{tab:mockdata}
\end{table*}

In this section, we present our joint tomographic BAO and RSD measurement from the BOSS DR12 combined sample, after validating our pipeline using the Patchy mock catalogue. We then perform a cosmological implication of our BAO and RSD measurement on the $f(R)$ gravity, and summarise the result.

\subsection{Pipeline validation using the Patchy mocks}
\label{sec:mocktest}

To validate our pipeline, we perform a joint BAO and RSD analysis on the Patchy mock catalogues, which are also subdivided into nine overlapping redshift slices detailed in Table \ref{tab:zbins}. We measure the BAO and RSD parameters (with other relevant parameters marginalised over) from the average of the power spectra multipoles derived from $2048$ Patchy mocks, and show the result in Table \ref{tab:mockdata}. As this is a mock test, which is used to validate our pipeline by checking whether we can reproduce the values of cosmological parameters used to create the mocks, we show the absolute values of the difference (multiplied by $100$ for the ease of visulisation) between the measurements and the values expected. 

To check the agreement between our measurement and expected values, we further define two quantities, namely the fractional bias $\Delta p/p$, and the fractional increase in the total uncertainty $\Delta \sigma_p/\sigma_p$,
\begin{flalign}
\label{eq:bias} &\Delta p/p \equiv \max_{\forall z_i \in Z_9} \  \left|p_{z_i}/p_{z_i,{\rm fid}}-1\right|,& \\
\label{eq:error}  &\Delta \sigma_p/\sigma_p \equiv  \max_{\forall z_i \in Z_9}  \left[\left(p_{z_i}-p_{z_i,{\rm fid}}\right)^2/\sigma_{p,{z_i}}^2+1\right]^{1/2}-1.&
\end{flalign}
where the set $p$ includes BAO and RSD parameters, \ie, $p\equiv\left\{\alpha_{\parallel},\alpha_{\perp},f\sigma_8\right\}$, and the set $Z_9$ is a collection of nine effective redshifts for this analysis, as shown in Table \ref{tab:zbins}. 

As defined, $\Delta p/p$ is the fractional measurement bias of the parameters, while $\Delta \sigma_p/\sigma_p$ quantifies the fractional increase in the total uncertainty of parameters due to the bias in the measurement. Note that we estimate the total uncertainty of a parameter by adding the bias and statistical uncertainty in quadrature, \ie, \be \sigma_p = \sqrt{\left(p-p_{\rm fid}\right)^2+\sigma^2_{\rm stat}} \ee Note that $\Delta p/p$ and $\Delta \sigma_p/\sigma_p$ are maximal values of the fractional measurement bias, and of the fractional increase in the total uncertainty respectively, across all the redshift slices.

We list $\Delta p/p$ and $\Delta \sigma_p/\sigma_p$ derived from monopole and quadrupole $(P_0+P_2)$, and from monopole, quadrupole and hexadecapole $(P_0+P_2+P_4)$ in Table \ref{tab:mocktest}. As shown, using $P_0+P_2$, the bias for $f\sigma_8$ is around $8\%$, although a higher level of agreement is reached for $\alpha_{\parallel}$ and $\alpha_{\perp}$, namely, the biases for these parameters never exceed $2\%$ in all redshift slices. However, the biases can dilute the total uncertainties by up to 17\%. Adding hexadecapole to the analysis, however, significantly reduces the bias, which avoids inflating the total error budget to a noticeable amount \citep{Beutler2017}. Specifically, the biases of the BAO parameters are reduced to sub-percent level, and the bias of $f\sigma_8$ drops by more than a factor of $2$. More importantly, $\Delta \sigma_p/\sigma_p$ is pushed below $8\%$ in all cases, which means that the systematic error budget is negligible compared with the statistical errors.

This mock test validates our pipeline, as we successfully recover the BAO and RSD parameters from the average of mock catalogues, with a negligible bias and impact on the total uncertainty when monopole, quadrupole and hexadecapole are used for the analysis.

\begin{table}
\caption{The fractional bias $\Delta p/p$ defined in Eq (\ref{eq:bias}) and the fractional increase in the total uncertainty $\Delta \sigma_p/\sigma_p$ defined in Eq (\ref{eq:error}) of parameters $\alpha_{\parallel},\alpha_{\perp}$ and $f\sigma_8$ derived from $P_0+P_2$ (left part) and $P_0+P_2+P_4$ (right) respectively.}
\begin{center} 
\begin{tabular}{ccccc}
\hline\hline 
        &      \multicolumn{2}{c}{Mock catalogue  $\left(P_0+P_2\right)$}          & \multicolumn{2}{c}{Mock catalogue $\left(P_0+P_2+P_4\right)$}          \\ \hline
 & $\Delta p/p$       		& $\Delta \sigma_p/\sigma_p$  &    $\Delta p/p$       & $\Delta \sigma_p/\sigma_p$ \\ \hline   
 $\alpha_{\parallel}$      	&  $1.7\%$    &  $10.1\%$   &  $0.99\%$        	& $7.4\%$\\
 $\alpha_{\perp}$          	&   $1.8\%$   &  $16.9\%$    &   $0.70\%$    	& $6.6\%$\\
 $f\sigma_8$                 	&   $8.3\%$   &  $15.1\%$    &    $3.5\%$   	& $4.6\%$\\
\hline\hline
\end{tabular}
\end{center}
\label{tab:mocktest}
\end{table}

\begin{table*}
\caption{Measurements (mean with the 68\% CL uncertainty) of BAO and RSD parameters including $\alpha_{\perp}$, $\alpha_{\parallel}$ and $f\sigma_8$ using $P_0+P_2$ (left part of the table) and $P_0+P_2+P_4$ (right) derived from the BOSS DR12 catalogue at nine effective redshifts. The $\chi^2/\nu$ columns show the reduced $\chi^2$, where $\nu$ is the number of degrees of freedom.}
\begin{center} 
\begin{tabular}{ccccccccc}
\hline\hline 
        &      \multicolumn{4}{c}{DR12  $\left(P_0+P_2\right)$}              			    & \multicolumn{4}{c}{DR12 $\left(P_0+P_2+P_4\right)$}          \\ \hline
$z_{\rm eff}$  & $\alpha_{\parallel}$ & $\alpha_{\perp}$&   $f\sigma_8$ & $\chi^2/\nu$   & $\alpha_{\parallel}$ &  $\alpha_{\perp}$ &  $f\sigma_8$ & $\chi^2/\nu$   \\ \hline
0.31	&	$	0.990  \pm	0.037  $	&	$	1.022 	\pm	0.032 	$	&	$	0.437   \pm  0.054  $   & $ 56/45  $ &	$	1.000 	\pm	0.034 	$	&	$	 1.021   	\pm	0.032 	$	&	$	0.452   \pm 0.053  $ & $ 65/73  $	\\
0.36	&	$	0.965  \pm	0.042  $	&	$	1.005 	\pm	0.025 	$	&	$	0.442   \pm  0.057  $   & $ 59/45  $ &	$	0.984 	\pm	0.028 	$	&	$	 1.006     \pm	0.025 	$	&	$	0.450   \pm 0.054  $ & $ 73/73  $	\\
0.40	&	$	0.956  \pm	0.049  $	&	$	0.994 	\pm	0.033 	$	&   $	0.459   \pm  0.062  $   & $ 64/45  $ &	$	0.977 	\pm	0.033 	$	&	$	 0.996   	\pm	0.030 	$	&	$	0.461   \pm 0.055  $ & $ 80/73  $	\\
0.44	&	$	0.967  \pm	0.038  $	&	$	1.023 	\pm	0.035 	$	&	$	0.472   \pm  0.071  $   & $ 61/45  $ &	$	1.007 	\pm	0.027 	$	&	$	 1.024   	\pm	0.025 	$	&	$	0.480   \pm 0.052  $ & $ 90/73  $	\\
0.48	&	$	0.989  \pm	0.037  $	&	$	1.034 	\pm	0.041 	$	&	$	0.482   \pm  0.070  $   & $ 71/45  $ &	$	1.020 	\pm	0.026 	$	&	$	 1.045   	\pm	0.021 	$	&	$	0.469   \pm 0.052  $ & $ 83/73  $	\\
0.52	&	$	1.007  \pm	0.038  $	&	$	1.051 	\pm	0.024 	$	&	$	0.478   \pm  0.064  $   & $ 70/45  $ &	$	1.031 	\pm	0.027 	$	&	$	 1.048   	\pm	0.024 	$	&	$	0.483   \pm 0.042  $ & $ 84/73  $	\\
0.56	&	$	0.995  \pm	0.033  $	&	$	1.036 	\pm	0.022 	$	&	$	0.476   \pm  0.058  $   & $ 70/45  $ &	$	1.008 	\pm	0.026 	$	&	$	 1.032   	\pm	0.022 	$	&	$	0.471   \pm 0.045  $ & $ 72/73  $	\\
0.59	&	$	0.965  \pm	0.046  $	&	$	1.013 	\pm	0.020 	$	&	$	0.445   \pm  0.062  $   & $ 75/45  $ &	$	0.991 	\pm	0.024 	$	&	$	 1.010   	\pm	0.021 	$	&	$	0.435   \pm 0.042  $ & $ 75/73  $	\\
0.64	&	$	0.958  \pm	0.038  $	&	$	1.015 	\pm	0.019 	$	&	$	0.421   \pm  0.067  $   & $ 54/45  $ &	$	0.988 	\pm	0.025 	$	&	$	 1.014   	\pm	0.019 	$	&	$	0.426   \pm 0.046  $ & $ 69/73  $	\\

\hline\hline
\end{tabular}
\end{center}
\label{tab:galaxydata}
\end{table*}

\begin{table*}
\caption{Mean and the 68\% CL uncertainty on BAO and AP parameters derived from $P_0+P_2$ (left part of the table) and $P_0+P_2+P_4$ (right) using the BOSS DR12 catalogue at nine effective redshifts. The unit for $D_{\rm A}$ and $D_{\rm V}$ is Mpc, and is $\rm km\, s^{-1} Mpc^{-1}$ for $H$. $F_{\rm AP}$ is dimensionless.}
\begin{center} 
\begin{tabular}{ccccccccc}
\hline\hline 
                & 			 &   DR12 $\left(P_0+P_2\right)$        				 &                    &    			 &   & 	DR12 $\left(P_0+P_2+P_4\right)$			&		          &		            \\ \hline
$z_{\rm eff}$  & $D_{\rm A}\left({r_{\rm s}^{\rm fid}}/{r_{\rm s}}\right)$  &  $H \left({r_{\rm s}}/{r_{\rm s}^{\rm fid}}\right)$  & $D_{\rm V}\left({r_{\rm s}^{\rm fid}}/{r_{\rm s}}\right)$ &  $F_{\rm AP}$  & $D_{\rm A}\left({r_{\rm s}^{\rm fid}}/{r_{\rm s}}\right)$   &  $H \left({r_{\rm s}}/{r_{\rm s}^{\rm fid}}\right)$  & $D_{\rm V}\left({r_{\rm s}^{\rm fid}}/{r_{\rm s}}\right)$ &  $F_{\rm AP}$  \\ \hline
0.31	&	$	986  	\pm	31  $	&	$	80.5  	\pm	5.8  $	&	$	1250    \pm  33  $	&	$	0.349  	\pm	0.025  	$	&	$	992  	\pm	28  $	&	$	79.7   \pm 4.2   $	&	$	1198    \pm 32 $	&	$	0.348    \pm 0.023  	$	\\
0.36	&	$	1074  	\pm	27  $	&	$	85.2  	\pm	9.4  $	&	$	1394    \pm  31  $	&	$	0.415  	\pm	0.021  	$	&	$	1075 	\pm	26  $	&	$	83.5   \pm 8.9   $	&	$	1394    \pm 29 $	&	$	0.341    \pm 0.019   	$	\\
0.40	&	$	1139  	\pm	38  $	&	$	87.9  	\pm	7.2  $	&   $	1514    \pm  41  $	&	$	0.467  	\pm	0.042  	$	&	$	1141  	\pm	33  $	&	$	86.0   \pm 6.5   $	&	$	1514    \pm 37 $	&	$	0.468    \pm 0.039  	$	\\
0.44	&	$	1243  	\pm	32  $	&	$	89.0  	\pm	3.4  $	&	$	1681    \pm  35  $	&	$	0.531  	\pm	0.036  	$	&	$	1245  	\pm	30  $	&	$	85.5   \pm 2.7   $	&	$	1681    \pm 32 $	&	$	0.530    \pm 0.034  	$	\\
0.48	&	$	1330  	\pm	27  $	&	$	89.2  	\pm	4.0  $	&	$	1843    \pm  29  $	&	$	0.586  	\pm	0.025  	$	&	$	1331  	\pm	25  $	&	$	86.5   \pm 3.7   $	&	$	1843    \pm 29 $	&	$	0.508    \pm 0.023  	$	\\
0.52	&	$	1392  	\pm	32  $	&	$	89.6  	\pm	7.8  $	&	$	1983    \pm  34  $	&	$	0.632  	\pm	0.026  	$	&	$	1387  	\pm	31  $	&	$	87.5   \pm 6.8   $	&	$	1946    \pm 33 $	&	$	0.582    \pm 0.024  	$	\\
0.56	&	$	1419  	\pm	30  $	&	$	92.9  	\pm	7.9  $	&	$	2070    \pm  33  $	&	$	0.686  	\pm	0.024  	$	&	$	1413  	\pm	28  $	&	$	91.7   \pm 6.1   $	&	$	2013    \pm 32 $	&	$	0.687    \pm 0.023  	$	\\
0.59	&	$	1430  	\pm	29  $	&	$	97.6  	\pm	4.9  $	&	$	2108    \pm  31  $	&	$	0.740  	\pm	0.023  	$	&	$	1425  	\pm	28  $	&	$	95.0   \pm 3.2   $	&	$	2102    \pm 29 $	&	$	0.739    \pm 0.021  	$	\\
0.64	&	$	1482  	\pm	28  $	&	$	101.1  	\pm	3.4  $	&	$	2239    \pm  28  $	&	$	0.819  	\pm	0.022  	$	&	$	1479  	\pm	26  $	&	$	98.0   \pm 3.2   $	&	$	2201    \pm 27 $	&	$	0.713    \pm 0.020  	$	\\

\hline\hline
\end{tabular}
\end{center}
\label{tab:DaHDvFap}
\end{table*}

\subsection{BAO and RSD measurement from the DR12 combined sample}

\begin{figure*}
\centering
{\includegraphics[scale=0.9]{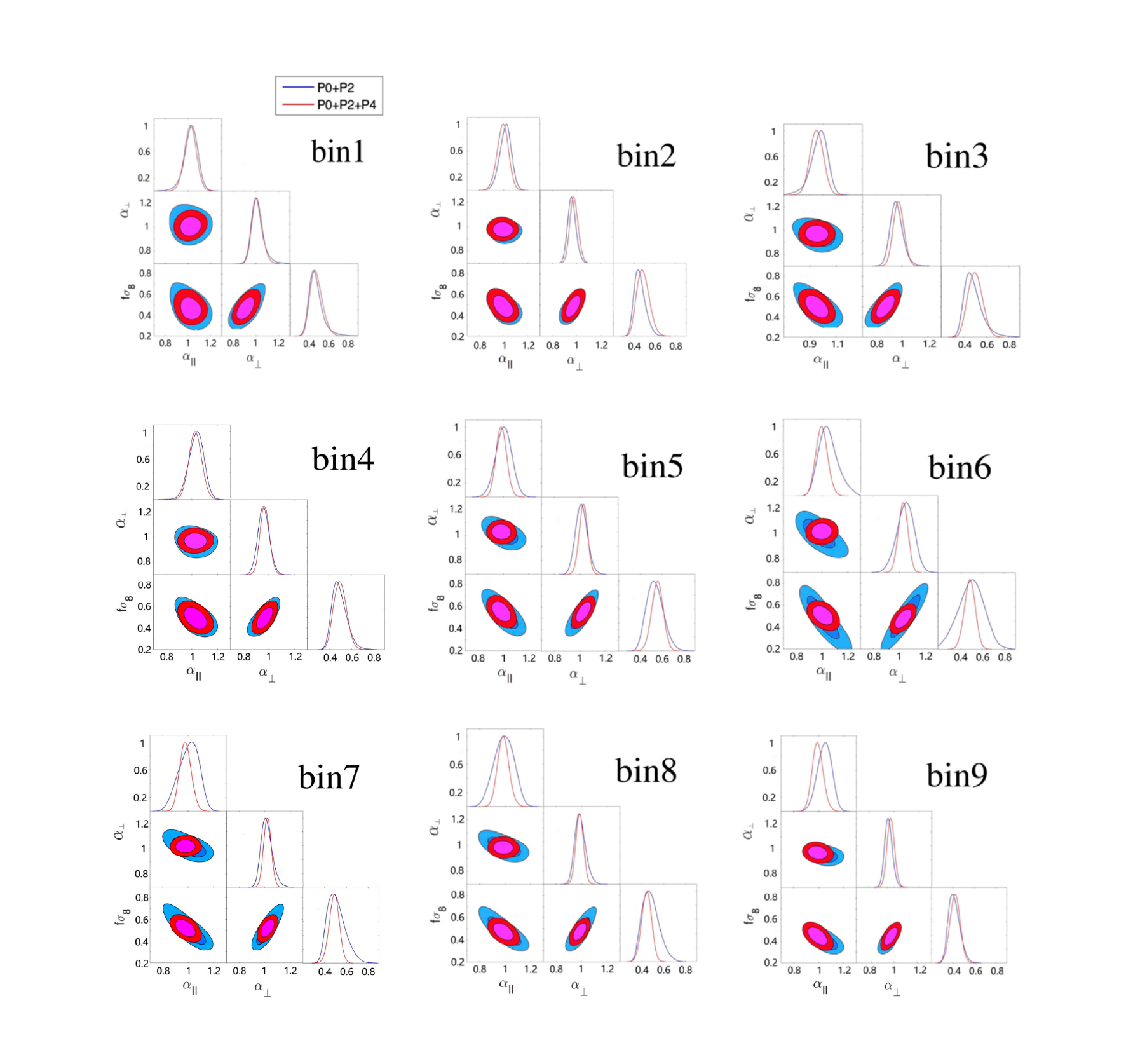}}
\caption{The one-dimensional posterior distribution and 68 and 95 \% CL contour plots for parameters $\alpha_{\perp}$, $\alpha_{\parallel}$ and $f\sigma_8$ derived from the BOSS DR12 catalogue at nine effective redshifts. The outer blue and inner red contours are derived from $P_0+P_2$ and $P_0+P_2+P_4$ respectively.}
\label{fig:zbin-contour}
\end{figure*}

\begin{figure*}
\centering
{\includegraphics[scale=0.33]{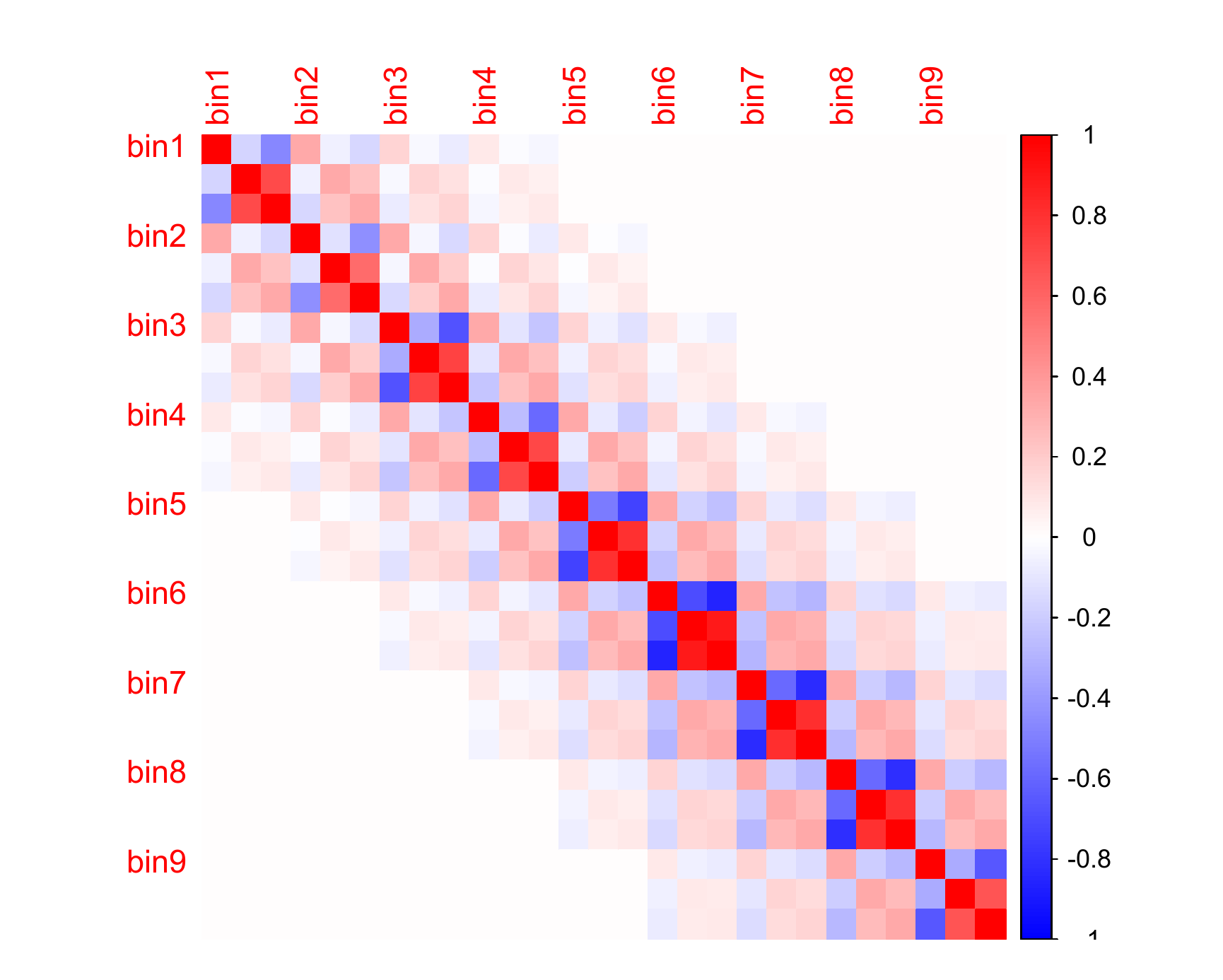}} {\includegraphics[scale=0.33]{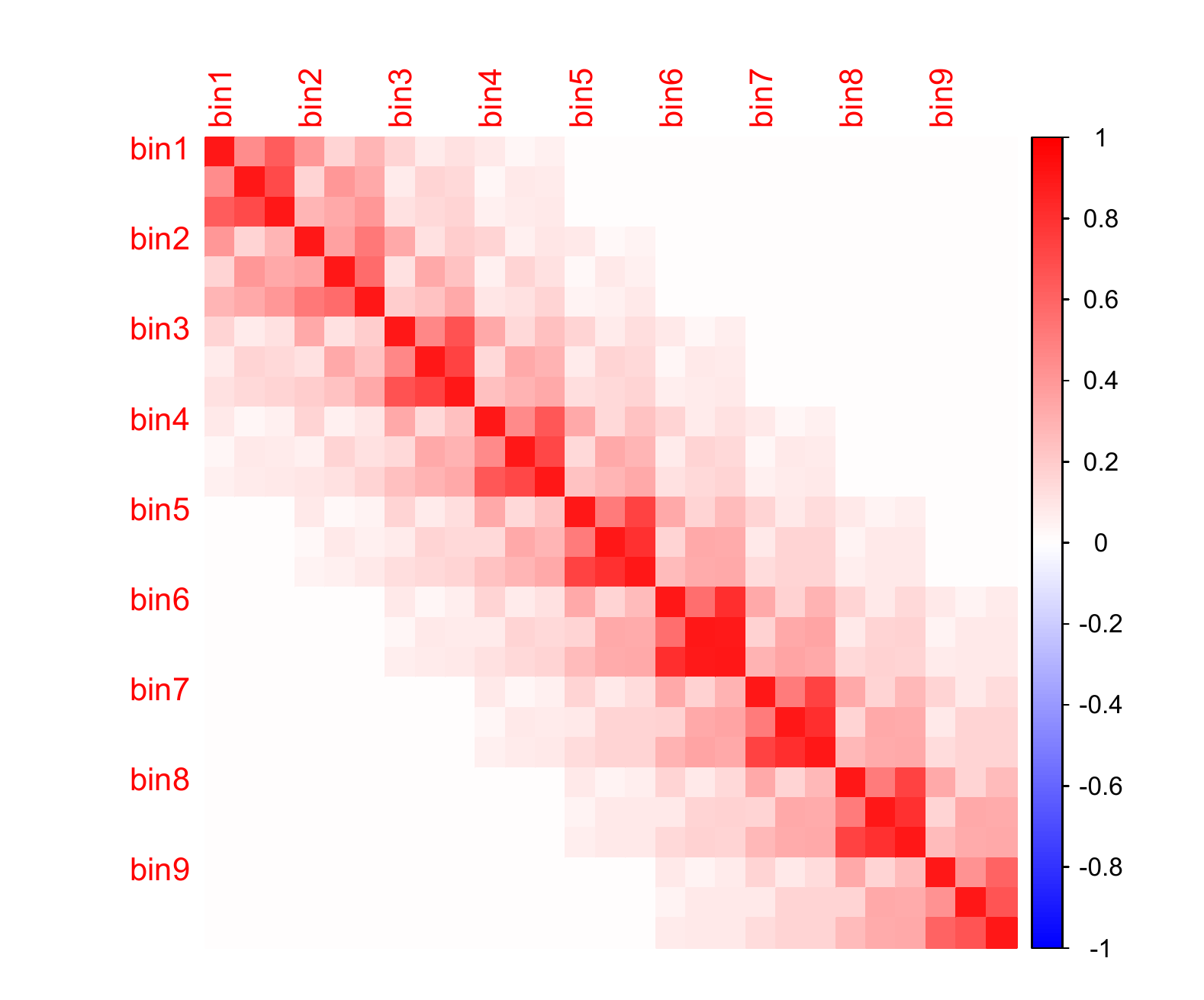}} {\includegraphics[scale=0.33]{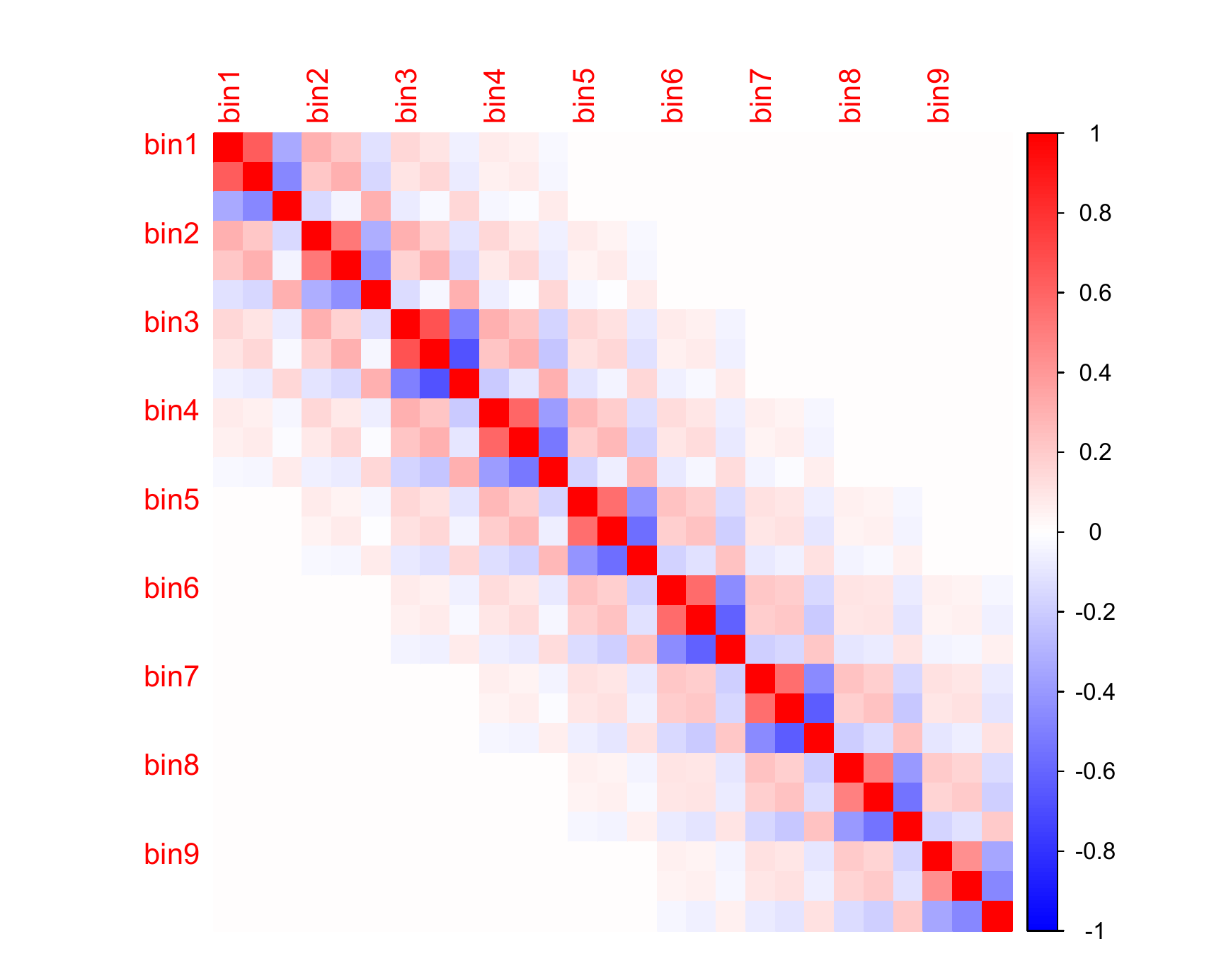}} 
\caption{The correlation matrix for parameters in order of \{$\alpha_{\parallel}, \alpha_{\perp}, f\sigma_8$\} (left panel),  \{$D_{\rm A}, H, f\sigma_8$\} (middle panel), and \{$D_{\rm V}, F_{\mathrm{AP}}, f\sigma_8$\} (right panel) at nine effective redshifts.}
\label{fig:big-corr}
\end{figure*}

\begin{figure}
\centering
{\includegraphics[scale=0.45]{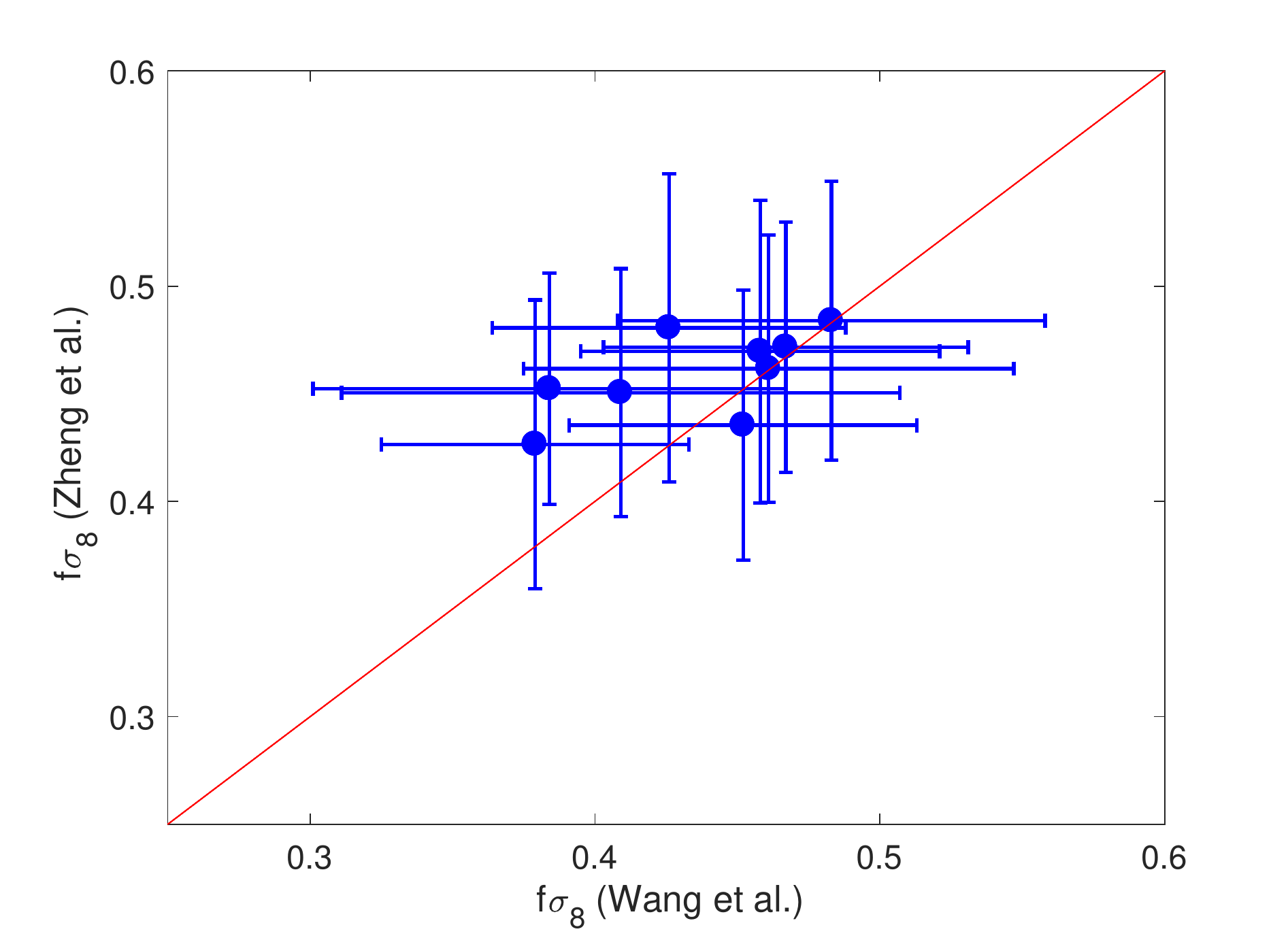}}
\caption{The mean and 68\% CL uncertainty of $f\sigma_8$ derived from this work (y-axis) in comparison with that in \citet{WangRSD} (x-axis) at nine effective redshifts.}
\label{fig:comparing}
\end{figure}

\begin{figure}
\centering
{\includegraphics[scale=0.33]{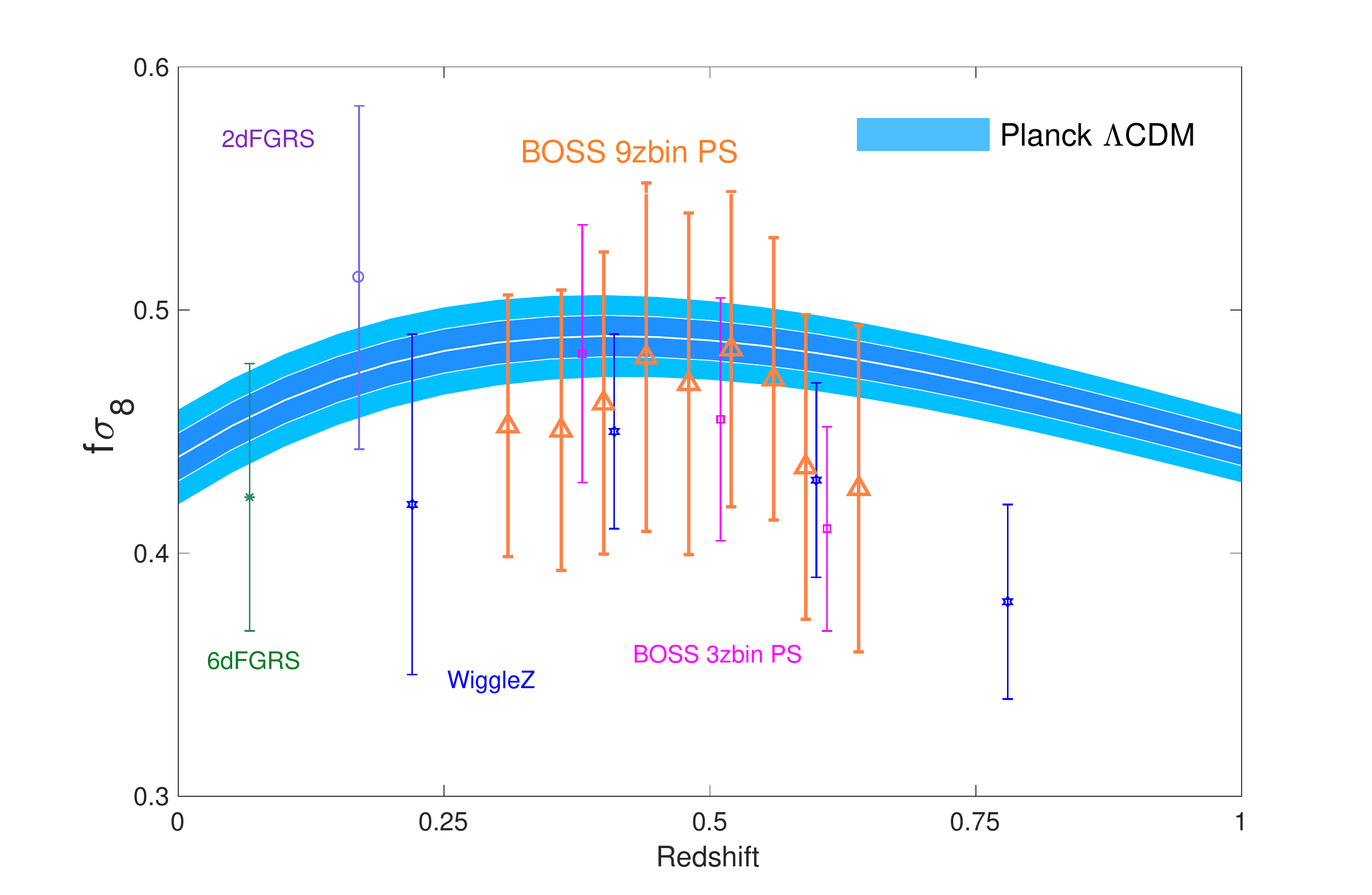}}
\caption{Measurements of $f\sigma_8$ in this work (denoted as `BOSS 9$z$bin PS') in comparison with those derived from the Planck data (deep and light blue bands indicating the 68 and 95\% CL uncertainties respectively) assuming a $\Lambda$CDM cosmology \citep{Planck2015}, 2dFGRS (a circle with error bar) \citep{Will2004}, 6dFGS (a star with error bar) \citep{Beutler2012}, BOSS at three effective redshifts (squares with error bars) \citep{Beutler2017} and WiggleZ (diamonds with error bars) \citep{Blake2011}.}
\label{fig:fs8Planckband}
\end{figure}

We apply our validated pipeline to the DR12 combined sample, and show the measurement of BAO and RSD parameters in Tables \ref{tab:galaxydata} and \ref{tab:DaHDvFap}, and in Figures \ref{fig:zbin-contour}, \ref{fig:big-corr}, \ref{fig:fs8Planckband} and \ref{fig:comparing}.

Table \ref{tab:galaxydata} and Figure \ref{fig:zbin-contour} show the constraints on $\alpha_{\parallel}, \alpha_{\perp}$ and $f\sigma_8$ derived from $P_0+P_2$ and $P_0+P_2+P_4$ respectively. As shown, $P_4$ can significantly improve the constraint on all three parameters, especially for redshift slices $5,6,7$ and $8$, in which $P_4$ is measured with relatively higher signal to noise ratio (SNR). Specifically, $\alpha_{\parallel}, \alpha_{\perp}$ and $f\sigma_8$ are measured with a precision of $3-5\%, 2-3\%$ and $12-16\%$ respectively, depending on the effective redshift, using $P_0+P_2$, and the precision is improved to $2-3\%, 2-3\%$ and $9-12\%$ when $P_4$ is added to the analysis. As demonstrated in the mock test in Sec. \ref{sec:mocktest}, analysis with $P_4$ included is not only more precise, but also more robust against systematics in the pipeline, we therefore regard our measurement derived from  $P_0+P_2+P_4$ as the main result of this work, and use it for comparison with other works and for cosmological implications.

We compare our $f\sigma_8$ measurement to that derived in \citet{WangRSD}, which applies the same overlapping redshift slicing scheme in configuration space. The comparison shown in Figure \ref{fig:comparing} demonstrates that our measurement is consistent with that in \citet{WangRSD} within 68\% CL at all redshifts. We also compare our measurement with that in \citet{Beutler2017}, which uses the same galaxy catalogue, but performs the measurement in three redshift slices. For comparison, we compress our measurement into that at three effective redshifts following the method in \citet{WangRSD}, and find an excellent agreement with \citet{Beutler2017}.

Figure \ref{fig:fs8Planckband} overplots our $f\sigma_8$ measurement with those derived from redshift surveys including BOSS DR12 \citep{Beutler2017}, WiggleZ~\citep{Blake2011}, 2dFGRS~\citep{Will2004} and 6dFGS~\citep{Beutler2012}. We also show the 68 and 95\% CL bands derived from the Planck mission assuming a $\Lambda$CDM model \citep{Planck2015}. As illustrated, our measurement enables a reconstruction of $f\sigma_8$ with high temporal resolution in the redshift range of $z\in[0.31,0.64]$, which provides key information for gravity tests.

For the ease of cosmological implications, we derive parameters related to physical BAO distances, including $D_A, H, D_V$ and $F_{\rm AP}$, from our measurement of $\alpha_{\perp}$ and $\alpha_{\parallel}$, where 
\begin{eqnarray}
Hr_s&\equiv&H^{\rm fid}r^{\rm fid}_s/\alpha_{\parallel}, \nn \\
\frac{D_A}{r_s}&\equiv&\alpha_{\perp}\frac{D^{\rm fid}_A}{r^{\rm fid}_s}, \nn \\
F_{\rm AP}&\equiv&\frac{\alpha_{\perp}}{\alpha_{\parallel}}(1+z)D^{\rm fid}_AH^{\rm fid}/c,\nn \\ 
\frac{D_V}{r_s}&\equiv&\left[\alpha_{\perp}^2\alpha_{\parallel}cz(1+z)^2D_{A,{\rm fid}}^2H^{-1}_{\rm fid}\right]^{1/3}.
\end{eqnarray} We show the result in Table \ref{tab:DaHDvFap}.

As our redshift slices overlap to a large extent, the parameters in each redshift slice correlate with those in other redshift slices. To obtain the data correlation matrix for BAO and RSD parameters, we perform joint fits of BAO and RSD parameters in each pair of redshift slices, and assemble. The resultant correlation matrices are shown in Figure \ref{fig:big-corr}. As expected, a positive correlation is seen for any given parameter with that in neighbouring redshift slices, and the correlation decays with separation of the slices in redshifts.

Our tomographic BAO and RSD measurement, including the data covariance matrix, is made publicly available \footnote{Our measurement is available at \url{https://github.com/Alice-Zheng/RSD-data}}. 

\subsection{A cosmological implication: constraining the $f(R)$ gravity}

In this section, we apply our tomographic BAO and RSD measurement to constrain one subset of modified gravity (MG) models, \ie, the $f(R)$ model.

Among various modified gravity models, the $f(R)$ gravity model has attracted much attention due to its simplicity (it is a one-parameter extension of general relativity), and its wide applicability (see \citealt{fR} for a review on the $f(R)$ gravity).

Generically, the effect of MG can be parametrised using two time- and scale-dependent functions $\mu(a,k)$ and $\eta(a,k)$, where $a$ and $k$ denote the scale factor and the wavenumber respectively, to modify the Poisson and anisotropic equations in the conformal Newton gauge,

\ba
k^2\Psi &=& 4\pi Ga^2\mu(a,k)\rho\Delta\;,\nn \\
\frac{\Phi}{\Psi}&=&\eta(a,k)\;,
\ea where $\Delta\equiv\rho\delta + 3\frac{aH}{k}(\rho+P)v$ denotes the comoving density perturbation. 

For general scalar-tensor theories, $\mu(a,k)$ and $\eta(a,k)$ can be parametrised as \citep{BZ}, \ba
\mu(a,k) &=& \frac{1+\beta_1 \lambda_1^2 k^2 a^s}{1+\lambda_1^2k^2a^s}, \nn \\
\eta(a,k) &=& \frac{1+\beta_2 \lambda_2^2 k^2 a^s}{1+\lambda_2^2k^2a^s},
\ea where $\beta_1$ and $\beta_2$ (denoting the coupling; dimensionless), $s$ (the power index; dimensionless), $\lambda_1$ and $\lambda_2$ (the length scales; in unit of Mpc) are free parameters. $a$ represents the scale factor.
In $f(R)$ theory, which is a special case of the scalar-tensor theory, the following relations hold,
\ba \beta_1=4/3; \  \ \ \beta_2=1/2; \ \ \ \lambda_2^2/\lambda_1^2=4/3. \ea
In addition, we fix $s=4$ to closely mimic the $\Lambda$CDM model at the background level \citep{AS09,1612Eva}. This only leaves one free parameter, $\lambda_1$, to be determined. In practice, we vary ${\rm log}_{10}B_0$ with other cosmological parameters where, 
 \ba
 B_0 \equiv \frac{2H_0^2 \lambda_1^2}{c^2}.
 \ea
The Hubble constant $H_0$ and the speed of light $c$ in the equation above make $B_0$ dimensionless, and the $\Lambda$CDM limit corresponds to $B_0=0$.

To constrain ${\rm log}_{10}B_0$, we use three different BAO and RSD measurements derived from the same DR12 combined galaxy sample, combined with the Planck 2015 observations \citep{Planck2015}, namely, 
\begin{itemize}
\item The consensus BAO and RSD measurement reported in \citet{Alam2017};
\item The tomographic BAO and RSD measurement in configuration space \citep{WangRSD};
\item This work.
\end{itemize}

\begin{figure}
\centering
{\includegraphics[scale=0.7]{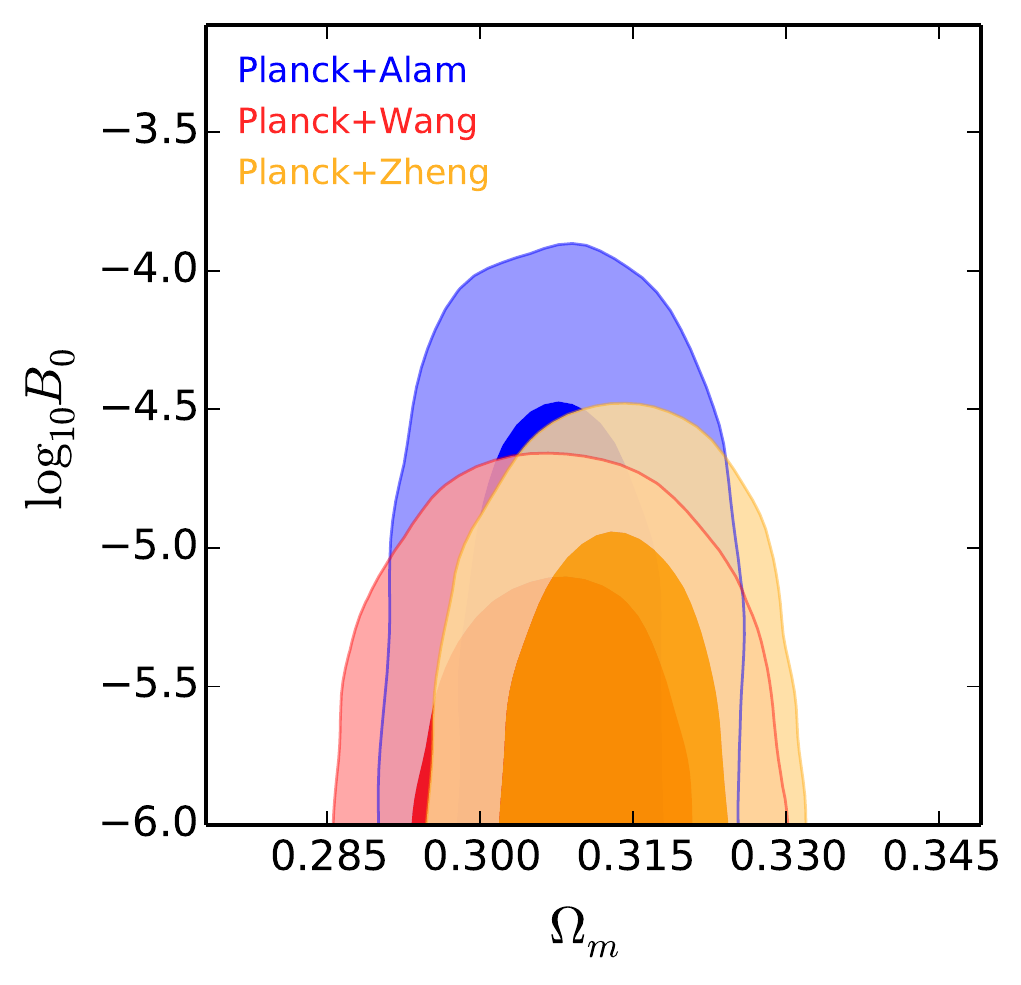}}
\caption{The 68 (inner) and 95\% (outer) CL contour plots for log$_{10}B_0$ and $\Omega_{\rm M}$ derived from Planck 15 combined with BAO and RSD measurements at three redshift slices (denoted as `Planck+Alam'), Planck 15 combined with BAO and RSD measurements in configuration space at nine redshift slices (denoted as `Planck+Wang'), and Planck 15 combined with BAO and RSD measurements in this work (denoted as `Planck+Zheng') respectively.}
\label{fig:MG}
\end{figure}

We use {\tt MGCAMB} \footnote{Available at \url{http://aliojjati.github.io/MGCAMB/}} \citep{MGCAMB1,MGCAMB2} and {\tt CosmoMC} for parameter estimation, and show the 68 and 95\% CL contour plot between ${\rm log}_{10}B_0$ and $\Omega_{\rm M}$ in Figure \ref{fig:MG}. As shown, the contours derived from `Planck + Wang' and `Planck + Zheng' (this work) are consistent with each other at 68\% CL, which is expected given the consistency between the BAO and RSD measurement in this work and that in \citet{WangRSD} (see Figure \ref{fig:comparing}). More importantly, the constraint on ${\rm log}_{10}B_0$ derived from `Planck + Alam' is much looser than that derived from `Planck + Wang' or `Planck + Zheng', namely,
\ba &&{\rm log}_{10}B_0 < -4.28 \ ({\rm Planck + Alam}) \nn \\ 
      &&{\rm log}_{10}B_0 < -4.76 \ ({\rm Planck + Zheng})\ea where the upper limit is for 95\% CL. This means that the tomographic information in `Planck + Zheng' reduces the upper limit of ${\rm log}_{10}B_0$ by 11\%, which is a nontrivial improvement in the constraint. This is expected, as we know that in $f(R)$ gravity, the enhancement of the growth due to $B_0$ varies with redshifts, thus tomographic measurements of the growth rate can help tighten the constraint of $B_0$. This demonstrates that our method successfully extracts additional information from the DR12 combined galaxy sample, which is able to tighten the constraint on MG parameters. A study on the observational constraint for a wide range of MG models using our tomographic BAO and RSD measurement with other latest observations is working in progress \citep{LZ18}.

\section{Conclusion and Discussions}
\label{sec:conclusion}

With the advance of large galaxy spectroscopic surveys, more and more information on the past light-cone becomes available for cosmological implications. In this work, we apply the overlapping redshift slicing method to the BOSS DR12 combined sample, and perform a joint tomographic BAO and RSD analysis in Fourier space, which largely complements \citet{WangRSD}, the analysis in the configuration space.

Splitting the BOSS DR12 galaxies into nine overlapping redshift slices, we obtain a joint measurement of $D_A, H$ and $f\sigma_8$ with a precision of $2-3\%, \ 3-10\%$ and $9-12\%$, depending on the effective redshifts, respectively. Our measurement covers the redshift range of $0.31<z<0.64$ with a redshift resolution as high as $\Delta z\sim0.04$. We apply our measurement to constrain the $f(R)$ gravity model for a proof-of-the-concept study, and find that the tomographic information extracted by our method improves the upper limit of the $f(R)$ model parameter ${\rm log_{10}}B_0$ by 11\%.

Efficient methods for extracting tomographic information from galaxy surveys have been actively developing \citep{GB2017,WangRSD,Wangtomo16,ZPW,zwBAOmock,zwRSD,RR18,DD18,Zhao18,Zhu18}, which have been proven advantageous for BOSS and eBOSS surveys. More efforts along this line, however, are needed for mitigating theoretical systematics in these methods for example, before making implications on upcoming deeper surveys such as DESI \footnote{More information is available at \url{http://desi.lbl.gov/}} \citep{DESI} and PFS \footnote{More information is available at \url{http://pfs.ipmu.jp/}} \citep{PFS}.

\section*{Acknowledgements} 

We are grateful to Shun Saito, who shared the TNS code with us, and commented on the manuscript. We also thank Florian Beutler and Will Percival for discussions.

JZ, GBZ, JL and YW are supported by the National Key Basic Research and Development Program of China (No. 2018YFA0404503), and by NSFC Grants 11720101004, 11673025 and 11711530207.

Funding for SDSS-III has been provided by the Alfred P. Sloan Foundation, the Participating Institutions, the National Science Foundation, and the US Department of Energy Office of Science. The SDSS-III web site is \url{http://www.sdss3.org/}. SDSS-III is managed by the Astrophysical Research Consortium for the Participating Institutions of the SDSS-III Collaboration including the University of Arizona, the Brazilian Participation Group, Brookhaven National Laboratory, Carnegie Mellon University, University of Florida, the French Participation Group, the German Participation Group, Harvard University, the Instituto de Astrofisica de Canarias, the Michigan State/Notre Dame/JINA Participation Group, Johns Hopkins University, Lawrence Berkeley National Laboratory, Max Planck Institute for Astrophysics, Max Planck Institute for Extraterrestrial Physics, New Mexico State University, New York University, Ohio State University, Pennsylvania State University, University of Portsmouth, Princeton University, the Spanish Participation Group, University of Tokyo, University of Utah, Vanderbilt University, University of Virginia, University of Washington, and Yale University.

This research used resources of the National Energy Research Scientific Computing Center, which is supported by the Office of Science of the U.S. Department of Energy under Contract No. DE-AC02-05CH11231, the SCIAMA cluster supported by University of Portsmouth.

\bibliographystyle{mnras} 
\bibliography{tomoRSDpk}

\bsp
\label{lastpage}

\end{document}